\documentclass[aip,rsi,reprint,graphicx]{revtex4-2} 

\usepackage{graphicx, color}
\usepackage{dcolumn}
\usepackage{bm}
\usepackage{here}
\usepackage{amsmath, amssymb}
\usepackage[whole]{bxcjkjatype}

\begin{document}

\title{Direct loading of a Sr magneto-optical trap from a thermal atomic beam}


\author{Naohiro Okamoto}
\author{Takumi Sato}
\author{Takatoshi Aoki}
\author{Yoshio Torii}
\email{ytorii@phys.c.u-tokyo.ac.jp}

\affiliation{Institute of Physics, The University of Tokyo, 3-8-1 Komaba, Meguro-ku, Tokyo 153-8902, Japan}
\date{\today}


\begin{abstract}
We demonstrate direct loading of a strontium (Sr) magneto-optical trap (MOT) from a thermal atomic beam in a single-chamber vacuum system.
The MOT operates without a Zeeman slower,
a slowing laser, a two-dimensional MOT, or differential pumping, while the entire system is maintained in the ultra-high-vacuum regime by a single ion pump.
At an oven temperature of $395\,\mathrm{{}^\circ C}$, the MOT captures up to $10^{7}$ ${}^{88}\mathrm{Sr}$ atoms with a loading rate of $10^{7}\,\mathrm{atoms\,s^{-1}}$, while sustaining a background gas pressure of $1 \times 10^{-9} \,\mathrm{Torr}$. 
At this oven temperature, the MOT lifetime limited by collisions with background gas is $\sim 5 \,\mathrm{s}$, with the atom number primarily constrained by light-assisted two-body collisions. 
Eliminating differential pumping and precooling stages significantly reduces the system's size, weight, and power requirements, providing a robust and practical platform for field-deployable and spaceborne optical lattice clocks, as well as a variety of other applications requiring compact ultracold atom sources.

\end{abstract}


\maketitle

\section{Introduction}
The alkaline-earth(-like) atoms possess electronic structures characterized by long-lived metastable states and ultra-narrow optical transitions. These features offer diverse applications, including precision measurements~\cite{S.L.Campbell2017, E.Oelker2019, T.L.Nicholson2015, W.F.McGrew2018, S.M.Brewer2019, T.Bothwell2019, N.Nemitz2016, BACONcolab2021}, tests of special relativity~\cite{P.Delva2017}, probing of gravitational redshift~\cite{M.Takamoto2020, T.Bothwell2022, X.Zheng2023}, quantum simulation~\cite{S.Kolkowits2017}, quantum information~\cite{A.Daley2008, N.Schine2022, R.Tao2025}, detection of gravitational waves~\cite{S.Kolkowits2016,M.Abe2021}, and search for dark matter~\cite{M.Abe2021, T.Kobayashi2022}.
Among them, optical lattice clocks based on Sr have been extensively investigated as strong candidates for the redefinition of the SI second~\cite{N.Dimarcq2024}.

The magneto-optical trap (MOT) serves as the first stage for cooling neutral atoms to ultralow temperatures. 
For alkali-metal atoms, particularly Rb and Cs, the vapor pressure at room temperature is sufficiently high to enable loading the atoms into the MOT via thermal vapor, commonly referred to as a vapor cell MOT~\cite{C.Monroe1990}. 
In contrast, alkaline-earth(-like) atoms exhibit extremely low vapor pressures at room temperature, making this loading method impractical. 
An exception has been demonstrated with Sr, where the entire MOT chamber was heated to nearly $300\,\mathrm{{}^\circ C}$ to realize a vapor cell MOT~\cite{K.R.Vogel1999, T.Dinneen1999, X.Xu2003}. 
However, this approach inevitably induces light shifts due to blackbody radiation (BBR), which severely limits applications such as optical lattice clocks. 
Consequently, loading of the MOT for alkaline-earth(-like) atoms has conventionally relied on atomic beams generated in a chamber separated from the MOT chamber by differential pumping.

For precooling thermal atomic beams, it is common to employ a Zeeman slower~\cite{T.Kurosu1990, I.Courtillot2003, R.Hill2014, C.Feng2024} and/or a two-dimensional (2D) MOT~\cite{C.Feng2024, I.Nosske2017, M.Barbiero2020, H.Lee2025}. 
However, field-deployable and spaceborne optical lattice clocks~\cite{M.Takamoto2020, N.Poli2014, S.B.Koller2017, J.Grotti2018, S.Origlia2018, W.Bowden2019, N.Ohmae2021} demand stringent constraints on size, weight, and power (SWaP)~\cite{Y.B.Kale2022}. 
Approaches relying on a Zeeman slower or a 2D MOT are fundamentally incompatible with the SWaP requirements of such devices.

Recently, several studies have demonstrated the use of thermally generated strontium vapors produced by laser ablation of strontium oxide~\cite{O.Kock2016} or metallic strontium granules~\cite{C.Hsu2022} for loading the MOT. 
However, these experiments suffer from limited reproducibility and difficulties in maintaining ultra-high-vacuum (UHV) conditions ($\lesssim 1 \times 10^{-9} \,\mathrm{Torr}$).
A low-power microstructured atomic oven for alkaline-earth-like elements has also been demonstrated~\cite{J.Pick2025}; however, its fabrication is demanding, and the design is not widely adopted.

Direct loading of a grating MOT using a Sr dispenser without employing a Zeeman slower has been demonstrated~\cite{A.Sitaram2020}, but maintaining UHV conditions requires a differential pumping tube and the number of trapped ${}^{88}\mathrm{Sr}$ atoms is limited to approximately $10^6$.
Direct loading of a grating MOT from a Sr atomic oven has also been demonstrated~\cite{S.Bondza2022}; however, it requires a slowing laser beam and employs two pumps (an ion pump and a non-evaporable getter pump), and the number of trapped ${}^{88}\mathrm{Sr}$ atoms is also limited to about $10^6$.


In this paper, we further advance the simplification of Sr MOT loading. The MOT is directly loaded from a thermal atomic beam emitted by a simple and compact atomic oven, without a slowing laser. The entire vacuum chamber is maintained in the UHV regime by a single ion pump, i.e., without employing differential pumping or an additional getter pump. By implementing proper thermal management of the atomic source oven~\cite{M.Schioppo2012}, the MOT captures up to $10^7$ ${}^{88}\mathrm{Sr}$ atoms within one second at an oven temperature of $395\,\mathrm{{}^\circ C}$, while the background gas pressure remains in the UHV regime. 
The key to this achievement is that, at this oven temperature and atom number, the MOT lifetime limited by both background collisions and light-assisted two-body collisions exceeds $1\,\mathrm{s}$. 
Our MOT system features significantly reduced SWaP, providing a robust and practical platform for field-deployable and spaceborne optical lattice clocks, as well as for a variety of other applications requiring compact ultracold atom sources. 
In fact, our MOT system has been employed in a series of our studies investigating the complex decay processes in the Sr MOT~\cite{N.Okamoto2024, N.Okamoto2025a, N.Okamoto2025b}, demonstrating its practicality and reliability.


\section{Theory}
\label{theory}
\begin{figure}
	\begin{center}
		\includegraphics[width=85mm]{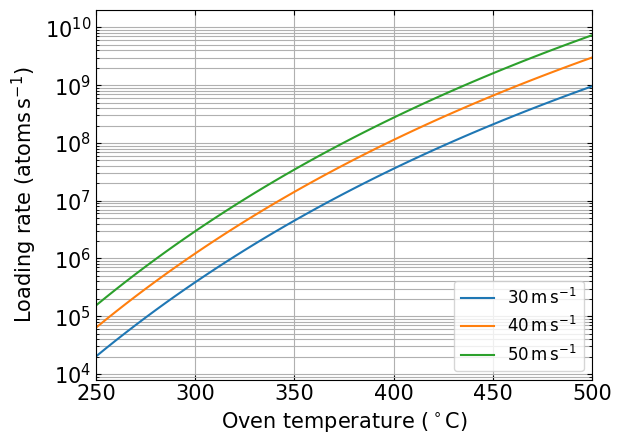}
		\caption{Theoretical curve of the MOT loading rate calculated for our vacuum system configuration and for various capture velocities.}
		\label{fig:theory_loading_rate}
	\end{center}
\end{figure}

\begin{figure*}
	\begin{center}
		\includegraphics[width=150mm]{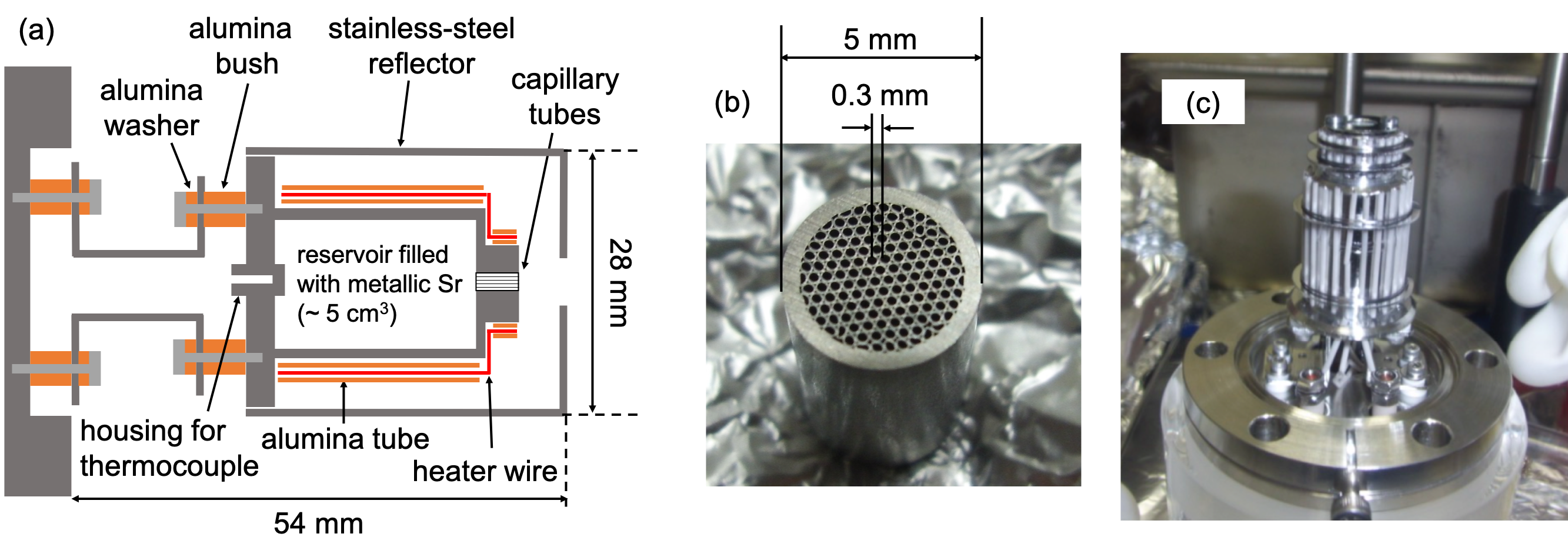}
		\caption{(a) Schematic illustration of the oven. Alumina washers and bushes are used to thermally isolate the oven from the vacuum system. (b) Picture of the capillary tubes. (c) Picture of the oven with the stainless-steel reflector removed.}
		\label{fig:oven}
	\end{center}
\end{figure*}

In the implementation of an optical lattice clock, the required number of atoms after the first-stage cooling at $461\,\mathrm{nm}$ is typically on the order of $10^7$. 
The atom number in the MOT is determined by its lifetime and loading rate. 
Since the interrogation time of an optical lattice clock is typically $1\,\mathrm{s}$, it is desirable that the MOT loading time be less than $1\,\mathrm{s}$. 
From these considerations, the MOT lifetime must exceed $1\,\mathrm{s}$ and the MOT loading rate must be greater than $10^7\,\mathrm{atoms\,s^{-1}}$.

We first discuss the MOT lifetime determined by collisions with background gas. 
Under UHV conditions, the background gas is predominantly composed of hydrogen ($\mathrm{H_2}$). 
The loss rate of the number of the trapped atoms due to collisions with $\mathrm{H_2}$ is calculated to be $5\times10^7\,\mathrm{s^{-1}\,Torr^{-1}}$ based on the model in Ref.~\cite{T.Arpornthip2012} (see Appendix~A). 
Therefore, obtaining an MOT lifetime longer than $1\,\mathrm{s}$ requires a background gas pressure lower than $2\times10^{-8}\,\mathrm{Torr}$.

Two-body collisions also limit the MOT lifetime. 
For Sr, the two-body loss coefficient $\beta$ (see Appendix~D) has been measured to be $4.5\times10^{-10}\,\mathrm{cm^3/s}$~\cite{T.Dinneen1999}. 
The typical diameter of the trap atomic cloud containing $10^7$ atoms is $\sim 2\,\mathrm{mm}$, corresponding to an atomic density of $10^9\,\mathrm{atoms\,cm^{-3}}$. 
From these, the two-body collisional loss rate for $10^7$ trapped atoms is estimated to be $\sim 0.5\,\mathrm{s^{-1}}$. 
Therefore, even if the effect of background gas collisions can be neglected, the MOT lifetime for $10^7$ atoms is limited to about $1\,\mathrm{s}$ by two-body collisions.

Next, we discuss the MOT loading rate. 
Following the method described in Ref.~\cite{M.Barbiero2020}, we estimated the capture velocity by simulation using our experimental parameters (see Sec.~\ref{Experimental setup}), obtaining a value of $\sim 50\,\mathrm{m\,s^{-1}}$. 
Meanwhile, the atomic beam flux from an oven can be derived from the geometry of the capillary tubes and the empirical expression for the vapor pressure (see Appendix~B). 
From these considerations, the MOT loading rate calculated for our vacuum system configuration is shown in Fig.~\ref{fig:theory_loading_rate} (see Appendix~C). 
The figure indicates that even if the capture velocity is $30\,\mathrm{m\,s^{-1}}$, an oven temperature of $\sim 400\,\mathrm{{}^\circ C}$ is sufficient to achieve a MOT loading rate of $10^7\,\mathrm{atoms\,s^{-1}}$.

As demonstrated by our experiment, proper thermal management of the oven allows the background gas pressure in the vacuum system to be maintained in the UHV regime at oven temperatures below $400\,\mathrm{{}^\circ C}$. 
Therefore, even without precooling of the atomic beam by a Zeeman slower or a 2D MOT, it is theoretically expected that $10^7$ atoms can be trapped in the MOT within one second.

\section{Experimental setup}
\label{Experimental setup}
\begin{figure}
	\begin{center}
		\includegraphics[width=85mm]{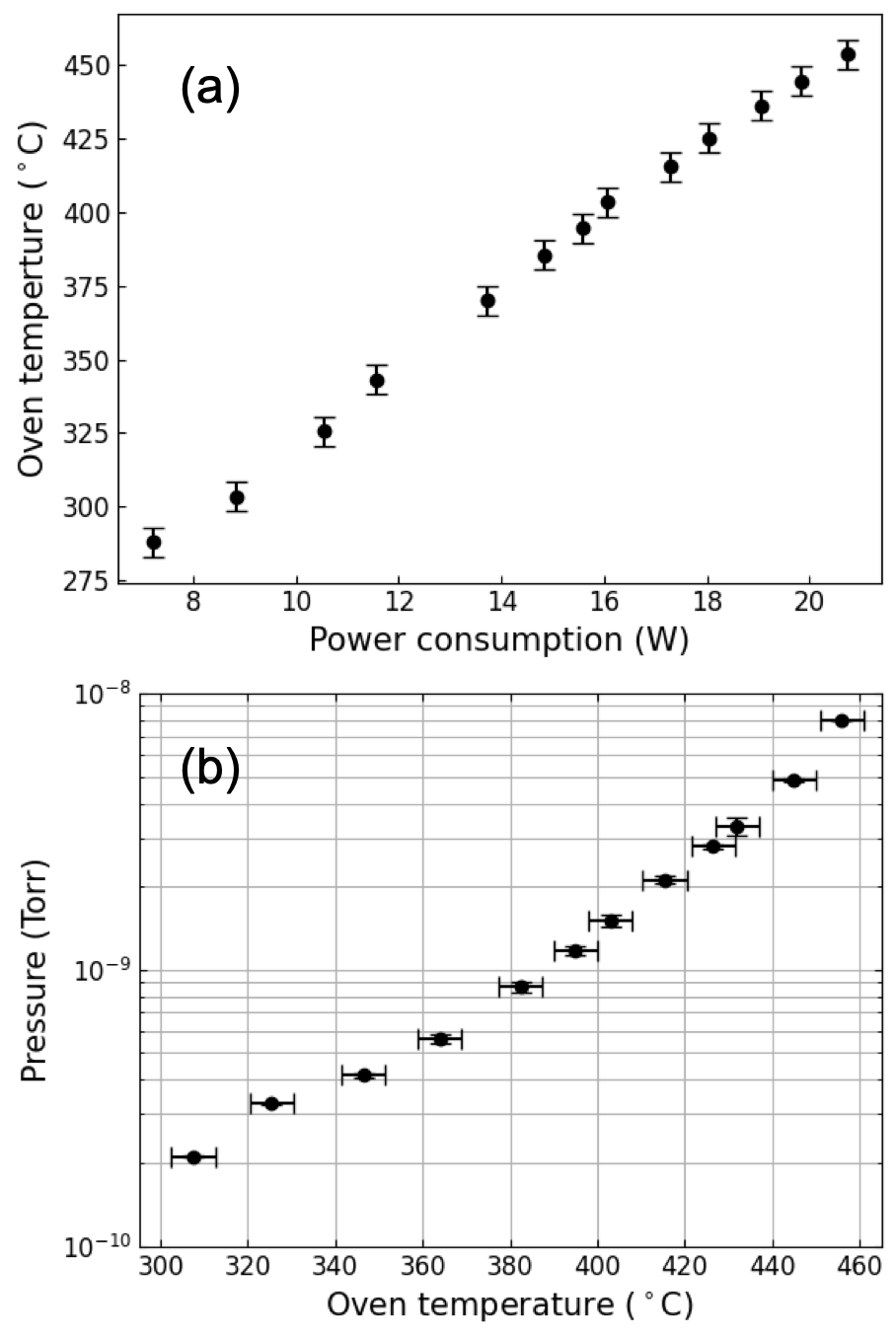}
		\caption{(a) Oven temperature as a function of power consumption. (b) Vacuum pressure as a function of oven temperature.}
		\label{fig:power_consumption}
	\end{center}
\end{figure}

\begin{figure}
	\begin{center}
		\includegraphics[width=85mm]{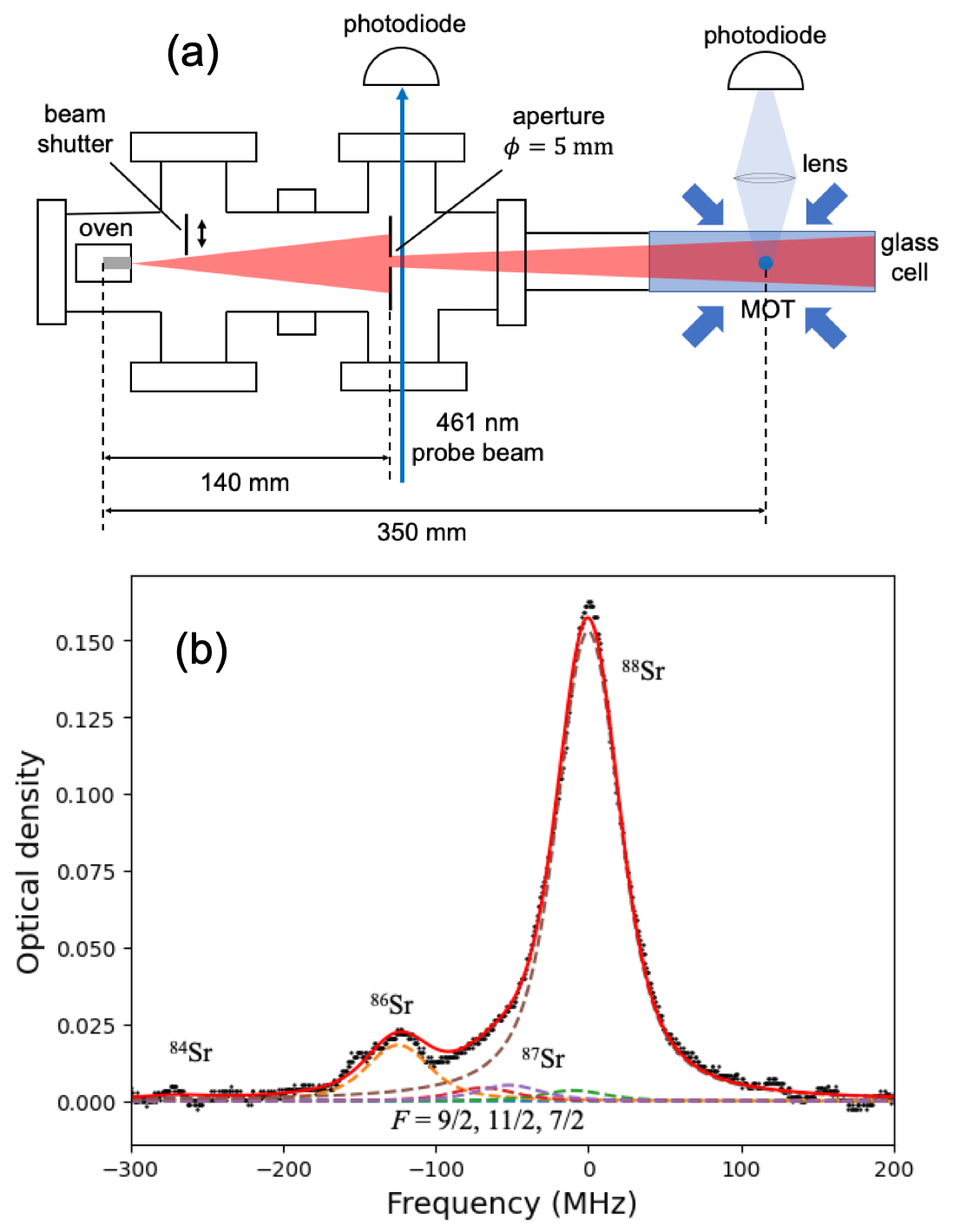}
		\caption{(a) Experimental setup for measuring the atomic beam flux and the number of trapped atoms. (b) Observed optical density (black dots) along the direction transverse to the atomic beam propagation at an oven temperature of $455\,\mathrm{{}^\circ C}$. The fit function (red line) is the sum of six Voigt functions [Eq.~\eqref{eq:OD_element}]. Each dashed line represents the contribution of the three bosonic isotopes $^{84}$Sr (0.56\%), $^{86}$Sr (9.86\%), and $^{88}$Sr (82.58\%), and the three hyperfine components of the fermionic $^{87}$Sr (7.00\% in natural abundance).}
		\label{fig:absorption}
	\end{center}
\end{figure}

\begin{figure}
	\begin{center}
		\includegraphics[width=85mm]{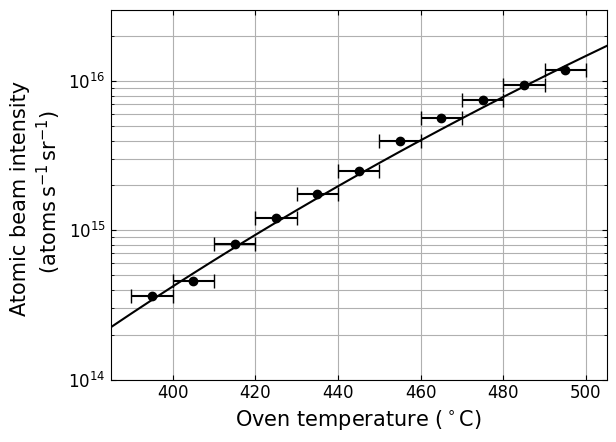}
		\caption{Measured atomic beam intensity as a function of the oven temperature, in comparison with the theoretical curve (Knudsen regime).}
		\label{fig:apparatus_results}
	\end{center}
\end{figure}


We first describe the dimensions and features of the vacuum apparatus. 
As shown in Fig.~\ref{fig:absorption}(a), the vacuum chamber is composed of two ICF70 six-way crosses, one of which is equipped with a glass cell ($30\,\mathrm{mm} \times 30\,\mathrm{mm} \times 100\,\mathrm{mm}$), an ion gauge, and an ion pump and the other with an oven and an atomic beam shutter.
A key advantage of this system is that UHV conditions are achieved using only a single ion pump (Varian VacIon Plus 55, pumping speed $55\,\mathrm{L\,s^{-1}}$), without the need for a differential pumping tube. 
The vacuum chamber fits within a volume of $500\,\mathrm{mm}\times 500\,\mathrm{mm}\times 150\,\mathrm{mm}$. 

The atomic beam is generated by heating a compact oven, as schematically illustrated in Fig.~\ref{fig:oven}, following the design reported in Ref.~\cite{M.Schioppo2012}. 
The beam is collimated using capillary tubes (made of SUS304) with an inner diameter of $0.3\,\mathrm{mm}$, an outer diameter of $0.4\,\mathrm{mm}$, and a length of $10\,\mathrm{mm}$, with 130 capillary tubes installed in total.
The oven contains $4\,\mathrm{g}$ of Sr, which is installed in the reservoir with the capillary tubes removed under an argon atmosphere.
Using Eq.~\eqref{eq:flowrate} in Appendix~B, we estimate the lifetime of the oven to be about 20 years at a temperature of $400\,\mathrm{{}^\circ C}$.
The temperature of the oven is measured by a thermocouple at the rear side of the reservoir. The oven is heated by a tantalum wire (diameter of $0.3\,\mathrm{mm}$) insulated by alumina tubes.
Thermal isolation from the vacuum chamber using alumina bushes and washers and radiation shielding using a stainless-steel cylindrical reflector enable high oven temperatures with reduced power consumption (Fig.~\ref{fig:oven}). 
For example, an oven temperature of $400\,\mathrm{{}^\circ C}$ requires only $16\,\mathrm{W}$ ($1.6\,\mathrm{A}, 10\,\mathrm{V}$) of heating power [Fig.~\ref{fig:power_consumption}(a)]. 
Figure~\ref{fig:power_consumption}(b) represents the dependence of the vacuum pressure measured by the ion gauge on the oven temperature.
The vacuum pressure remains in the UHV regime ($\lesssim 1\times10^{-9}\,\mathrm{Torr}$) at temperatures below $380\,\mathrm{{}^\circ C}$, presumably owing to the getter effect of metallic Sr deposited on the surfaces of the vacuum chamber and the stainless-steel reflector.

The atomic beam passes through a $5\,\mathrm{mm}$-diameter aperture, which is positioned $140\,\mathrm{mm}$ downstream from the oven to prevent Sr deposition on the side walls of the glass cell and thereby maintain the transmission of the trapping light. After five years of oven operation, no visible Sr deposition on the side walls of the glass cell has been observed, while the inner surface at the end of the glass cell is coated with metallic Sr.
 
To determine the atomic beam flux, we shine a probe beam at $461\,\mathrm{nm}$ orthogonal to the atomic beam axis just after the aperture [Fig.~\ref{fig:absorption} (a)].
Figure~\ref{fig:absorption} (b) shows the optical density obtained from the absorption spectrum at an oven temperature of $455\,\mathrm{{}^\circ C}$. 
The atomic beam intensity derived from this optical density is shown in Fig.~\ref{fig:apparatus_results} (see Appendix~B). 
The agreement between the theoretical and experimental values shown in Fig.~\ref{fig:apparatus_results} indicates that the experiment was conducted in the Knudsen regime, where no collisions occur inside the capillary tubes.

As shown in Fig.~\ref{fig:absorption} (a), the MOT is implemented for ${}^{88}\mathrm{Sr}$ inside the glass cell.
The distance from the oven to the MOT region is $350\,\mathrm{mm}$. 
The MOT is operated using three laser wavelengths: $461\,\mathrm{nm}$ (the $5s^2 \,{}^1S_0 - 5s5p \,{}^1P_1$ transition for trapping), $481\,\mathrm{nm}$ (the $5s5p \,{}^3P_2 - 5p^2 \,{}^3P_2$ transition for ${}^3P_2$ repumping), and $483\,\mathrm{nm}$ (the $5s5p \,{}^3P_0 - 5s5d \,{}^3D_1$ transition for ${}^3P_0$ repumping). 
Experimental parameters are as follows: axial magnetic field gradient of $55\,\mathrm{G\,cm^{-1}}$, trapping light detuning frequency of $-40\,\mathrm{MHz}$, beam diameter of $18\,\mathrm{mm}$, and peak intensity of $65\,\mathrm{mW\,cm^{-2}}$.
The number of trapped atoms is measured by detecting atom fluorescence using a photodiode.
Further details can be found in Ref.~\cite{N.Okamoto2024}.

\section{Results and Discussion}
\begin{figure}
	\begin{center}
		\includegraphics[width=85mm]{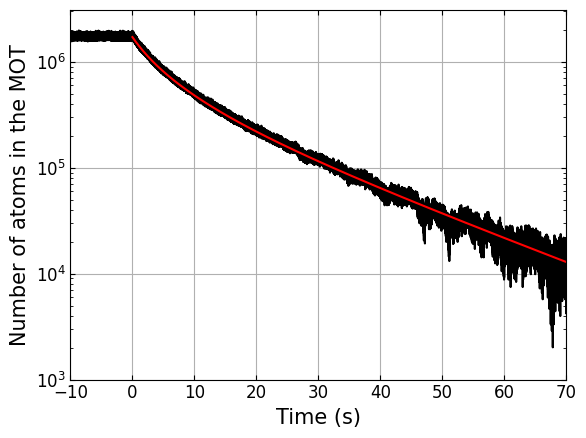}
		\caption{Decay curve of the trapped atom number at an oven temperature of $325\,\mathrm{{}^\circ C}$. An initial rapid decay due to two-body collisions is observed, followed by a slower decay with a longer time constant attributed to collisions with background gas. A red curve is a fit to the data using a solution of Eq.~\eqref{eq:rate_general} with $R=0$.}
		\label{fig:MOT_decay}
	\end{center}
\end{figure}

\begin{figure}
	\begin{center}
		\includegraphics[width=85mm]{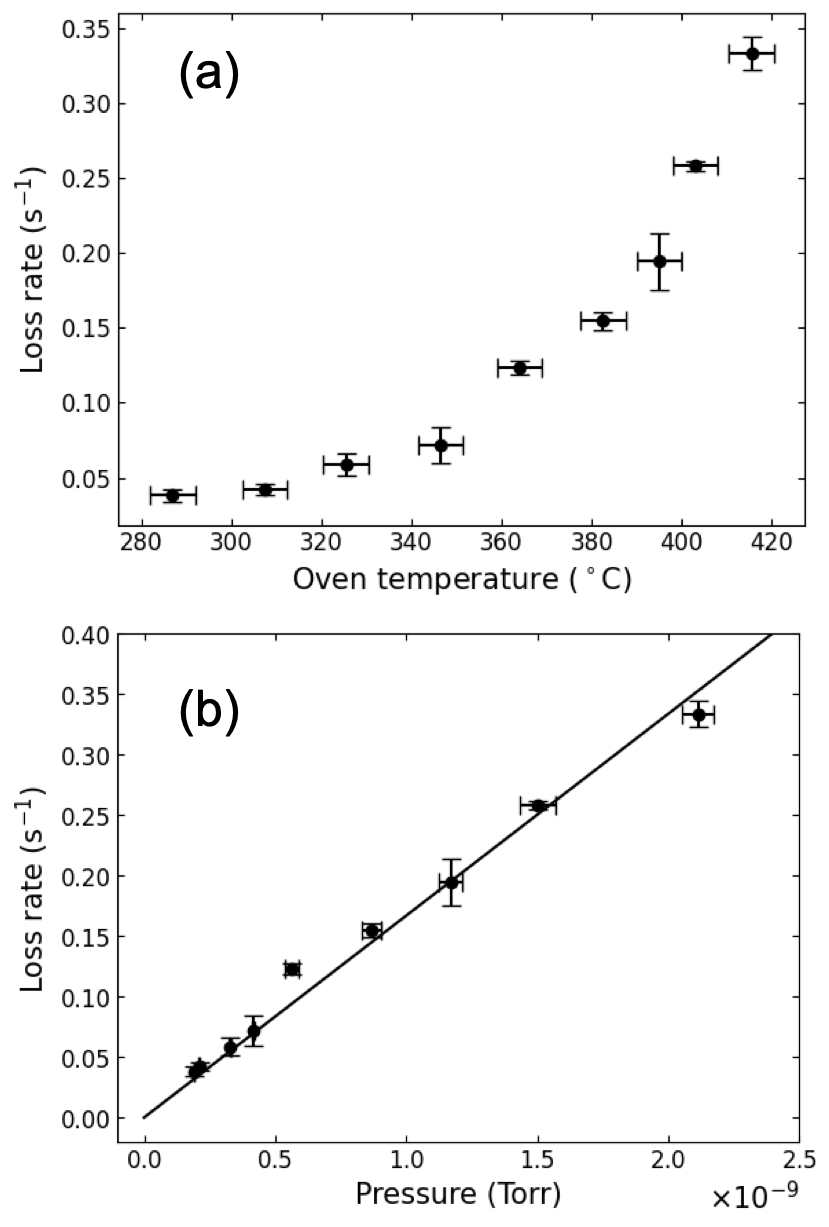}
		\caption{(a) Temperature dependence of the loss rate due to background gas collisions. (b) Loss rate due to background gas collisions versus background gas pressure measured by the ion gauge. The solid line is a linear fit.}
		\label{fig:lifetime_pressure}
	\end{center}
\end{figure}

\begin{figure}
	\begin{center}
		\includegraphics[width=85mm]{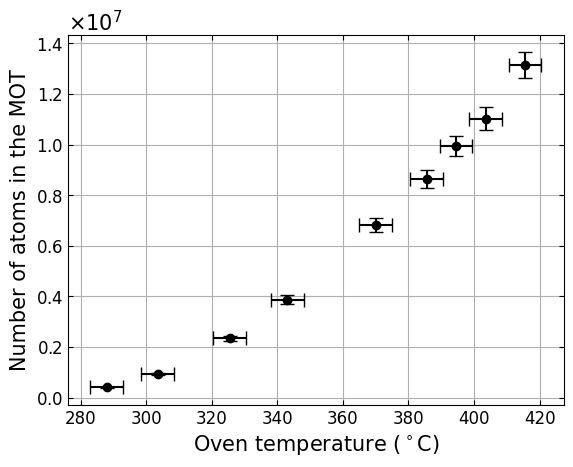}
		\caption{Trapped atom number as a function of oven temperature.}
		\label{fig:atomnumber}
	\end{center}
\end{figure}


\begin{figure}
	\begin{center}
		\includegraphics[width=85mm]{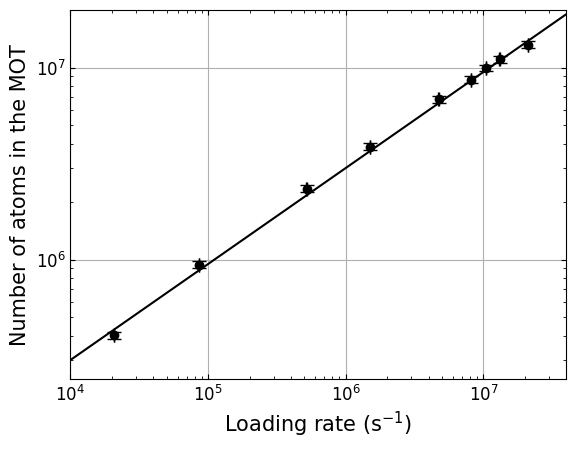}
		\caption{Trapped atom number versus loading rate. The solid line represents a fit assuming a square-root dependence on the loading rate, yielding an effective two-body loss coefficient of $\tilde{\beta}=1.11(3) \times 10^{-7}\,\mathrm{s^{-1}}$ (see Appendix~D).}
		\label{fig:loading_dual}
	\end{center}
\end{figure}

To estimate the loss rate of trapped atoms due to collisions with background gas, we block the atomic beam with the beam shutter and monitor the decay curve of the number of trapped atoms.
Figure~\ref{fig:MOT_decay} shows the decay curve of the trapped atom number at an oven temperature of $325\,\mathrm{{}^\circ C}$. 
The initial fast decay is due to two-body collisions, which is followed by the slow exponential decay caused by collisions with background gas.
Figure~\ref{fig:lifetime_pressure}(a) shows the dependence of the loss rate due to collisions with background gas on the oven temperature, obtained by fitting the latter decay with an exponential function. 
As seen in the figure, this loss rate increases with oven temperature. 
Figure~\ref{fig:lifetime_pressure}(b) demonstrates a linear relationship between background gas pressure measured by the ion gauge and the loss rate, with a coefficient of $\alpha_{\text{fit}} = 1.67(4) \times 10^8\,\mathrm{s^{-1}\,Torr^{-1}}$. 
In contrast, the coefficient calculated using Arpornthip's expression~\cite{T.Arpornthip2012} and the $C_6$ coefficients from Ref.~\cite{M.Abdel-Hafiz2019} is $\alpha = 5.2(5) \times 10^7\,\mathrm{s^{-1}\,Torr^{-1}}$ (see Appendix~A), giving a ratio of $\alpha_{\text{fit}}/\alpha = 3.2(3)$. 
This indicates that the background gas pressure in the glass cell is approximately three times higher than that measured by the ion gauge. 
A plausible explanation is a poor conductance ($\sim 5\,\mathrm{L\,s^{-1}}$) of the neck of the glass cell.

Figure~\ref{fig:atomnumber} shows the dependence of the trapped atom number on the oven temperature. 
It can be seen that around $400\,\mathrm{{}^\circ C}$, where the background gas pressure is in the range of $1\times10^{-9}\,\mathrm{Torr}$, up to $10^7$ atoms are successfully trapped, consistent with the theoretical estimation presented in Sec.~\ref{theory}.

Finally, we demonstrate that, for temperatures below $420\,\mathrm{{}^\circ C}$, the trapped atom number is limited by two-body collisions.
Figure~\ref{fig:loading_dual} shows the dependence of trapped atom number on the loading rate, which is determined by the initial slope of the loading curve.
The trapped atom number scales with the square root of the loading rate. 
This observation indicates that the trapped atom number is determined not by collisions with background gas, but rather by two-body collisions (see Appendix~D). 
This result demonstrates that at the temperatures below $420\,\mathrm{{}^\circ C}$, the vacuum system is maintained at a sufficiently low pressure such that two-body collisions dominate the limitation of the trapped atom number.
Note that even in the regime of low temperature ($\sim 300\,\mathrm{{}^\circ C}$) and low atomic density ($10^7-10^8\,\mathrm{cm^{-3}}$), the dominant loss mechanism of the MOT is two-body collisions.

Several approaches have been demonstrated to mitigate light-assisted two-body loss and increase the effective atom number in the MOT. One widely used approach is to accumulate atoms in a metastable ${}^3P_2$ magnetic-trap reservoir and subsequently recapture them into the MOT, which has been shown to enhance the effective atom number by one to two orders of magnitude~\cite{T.Mukaiyama2003, S.Stellmer2009, S.Stellmer2014}. Another approach is to apply additional $689\,\mathrm{nm}$ light to shelve atoms into the ${}^3P_1$ state during MOT loading, which reduces effective two-body loss and has been reported to increase the steady-state atom number by approximately a factor of two~\cite{J.Hoschele2023}.

In the present study, the performance of the MOT was evaluated using only $^{88}$Sr (82.58\% natural abundance). For bosonic isotopes without hyperfine structure, such as $^{86}$Sr (9.86\%) and $^{84}$Sr (0.56\%), the steady-state atom number is expected to follow the square-root scaling with loading rate discussed above. Since the loading rate in a thermal atomic beam is proportional to the isotopic abundance, the trapped atom number should therefore scale approximately with the square root of the natural isotopic abundance under otherwise identical conditions.

This simple scaling argument, however, may not directly apply to $^{87}$Sr (7.00\%), a fermionic isotope widely used in state-of-the-art optical lattice clocks.
Owing to its hyperfine structure, population redistribution among multiple magnetic sublevels and the corresponding reduction of effective transition strengths can reduce the effective scattering force in a MOT~\cite{T.Mukaiyama2003}, resulting in a reduced MOT capture velocity. Since the loading rate for MOTs directly loaded from thermal atomic beams scales with the fourth power of the capture velocity (Appendix~C), even a modest reduction in capture velocity can strongly suppress the loading rate, making the scheme less favorable for $^{87}$Sr. Consistent with this scenario, compact grating MOT systems directly loaded from dispenser-based thermal atomic beams have reported markedly reduced trapping of $^{87}$Sr~\cite{A.Sitaram2020, H.Lee2025}.

It is worth noting that transportable optical lattice clocks based on bosonic $^{88}$Sr have achieved high performance suitable for field deployment~\cite{S.Origlia2018}. This underscores the practical relevance of the present loading scheme despite the possible limitations for $^{87}$Sr.

\section{Conclusion}
We have realized a Sr MOT in a single-chamber vacuum system without employing a Zeeman slower, a slowing beam, a two-dimensional MOT, or differential pumping.
The MOT is directly loaded from a thermal atomic beam generated by a compact oven located 350 mm away.
By implementing proper thermal management of the oven and maintaining UHV conditions with a single ion pump, the system achieves efficient trapping of atoms without compromising the vacuum quality.
At an oven temperature of $395\,\mathrm{{}^\circ C}$, the MOT contained $10^7$ ${}^{88}\mathrm{Sr}$ atoms with a loading rate of $10^7\,\mathrm{atoms\,s^{-1}}$, consistent with the theoretical expectation.
The MOT lifetime is limited primarily by light-assisted two-body collisions rather than by background gas collisions, demonstrating that the vacuum conditions are sufficient to suppress background-gas collisional losses.
These results present a simple and robust design that significantly reduces system complexity, size, weight, and power consumption.
The single-chamber Sr MOT platform is suitable not only for field-deployable and spaceborne optical lattice clocks but also for other  applications such as quantum sensing, quantum information, and tests of fundamental physics.



\section{acknowledgments}
We thank Dr. H. Hachisu at the National Institute of Information and Communications Technology, Japan, for providing us with stainless-steel capillary tubes, and M. Mori for his contributions to the experiments.
This work was supported by JSPS KAKENHI Grant Numbers 23K20849 and 22KJ1163.

\appendix
\section{Dependence of the loss rate of trapped atoms on background gas pressure}
\label{appendix:background}
The rate equation for the atom number in the MOT in the absence of two-body collisions can be written as
\begin{equation}
    \frac{dN}{dt} = R - \gamma N, \label{eq:rate_background}
\end{equation}
where $R$ is the loading rate of trapped atoms and $\gamma$ denotes the loss coefficient due to collisions with background gas. 
The dominant component of the background gas is hydrogen ($\mathrm{H_2}$). 
In the following, we assume that the background gas consists solely of $\mathrm{H_2}$. 
Here, $\gamma$ is proportional to the partial pressure of hydrogen $P_{\mathrm{H_2}}$. 
If the long-range interaction potential between $\mathrm{Sr}$ and $\mathrm{H_2}$ is given by $-C_{\mathrm{Sr-H_2}}/r^6$ (where $r$ is the distance between $\mathrm{Sr}$ and $\mathrm{H_2}$), $\gamma$ can be expressed as~\cite{T.Arpornthip2012}
\begin{equation}
\gamma = \alpha P_{\mathrm{H_2}},
\label{eq:gamma_background}
\end{equation}
\begin{align}
\alpha &= \frac{6.8}{\left(k_\mathrm{B} T\right)^{2/3}}
\left(\frac{C_{\mathrm{Sr-H_2}}}{m_{\mathrm{H_2}}}\right)^{1/3}
\left(D m_{\mathrm{Sr}}\right)^{-1/6},
\label{eq:alpha_background}
\end{align}
where $k_\mathrm{B}$ is the Boltzmann constant, $T$ is the background gas temperature ($300\,\mathrm{K}$), $m_{\mathrm{H_2}}$ is the mass of $\mathrm{H_2}$, $m_{\mathrm{Sr}}$ is the mass of Sr, and $D$ is the MOT trap depth ($\sim 10 \,\mathrm{K}$). 
It is noteworthy from Eqs.~\eqref{eq:gamma_background} and \eqref{eq:alpha_background} that the background-gas-limited loss depends only weakly on the interaction potential $C_{\mathrm{Sr-H_2}}$, the MOT potential depth $D$, and the mass of the trapped atom $m_{\mathrm{Sr}}$.

Here, by using $C_{\mathrm{Sr-H_2}} = 166(17)\,E_{h}a_{0}^6$ from Ref.~\cite{M.Abdel-Hafiz2019}, 
where $E_h$ is the Hartree energy and $a_0$ is the Bohr radius, we obtain
\begin{equation}
    \alpha = 5.2(5) \times 10^7\,\mathrm{s^{-1}\,Torr^{-1}}.
\end{equation}

\section{Atomic beam flux}
We determine the atomic beam flux by measuring the absorption profile of the atomic beam emitted from the oven as shown in Fig.~\ref{fig:absorption}. 
The arguments presented in this section are primarily based on Ref.~\cite{M.Schioppo2012}.

To measure the atomic beam density, we evaluate the optical density (OD). 
The OD is defined as
\begin{equation}
\mathrm{OD}(\nu) = -\ln\!\left(\frac{I(\nu)}{I_0}\right),     \label{eq:OD_def}
\end{equation}
where $\nu$ is the laser frequency, $I_{\text{0}}$ is the incident laser intensity, and $I (\nu)$ is the transmitted intensity measured by a photodetector. 
The OD as a function of frequency $\nu$ can be expressed as a sum of Voigt profiles corresponding to individual isotopic and hyperfine components,

\begin{equation}
\mathrm{OD}(\nu) = \sum_i \mathrm{OD}_i(\nu),
\label{eq:OD_summation}
\end{equation}
\begin{equation}
\mathrm{OD}_{i}(\nu)
=
\frac{n r_{i} \sigma_{0} l_{\text{int}}}{\sqrt{2\pi}\,\sigma_{t}}
\int_{-\infty}^{\infty}
\frac{\exp\!\left(-\frac{v_t^2}{2\sigma_t^2}\right)}
{1 + 4\left(\frac{\nu-\nu_0^{i}+v_t/\lambda}{\Gamma/2\pi}\right)^2}
\, dv_t ,
\label{eq:OD_element}
\end{equation}
where $n$ is the atomic beam density, $r_i$ is the isotopic abundance of isotope $i$ for bosons (${}^{84}\mathrm{Sr}, {}^{86}\mathrm{Sr}, {}^{88}\mathrm{Sr}$) and the relative contribution of each hyperfine line $i$ for fermion (${}^{87}\mathrm{Sr}$), $\sigma_0 = 3\lambda^2/(2\pi)$ is the resonant scattering cross section without Doppler broadening, $l_{\text{int}}$ is the thickness of the atomic beam along the probe direction (set to $l_{\text{int}} = 5\,\mathrm{mm}$ from the aperture size), $\sigma_t$ is the transverse Doppler width, $v_t$ is the transverse atomic velocity, $\nu_0^{i}$ is the resonance frequency of isotope $i$ for bosons and of each hyperfine line $i$ for fermion, $\lambda$ is the probe wavelength ($461\,\mathrm{nm}$), and $\Gamma$ is the natural linewidth (FWHM) of the transition ($2\pi \times 30.2\,\mathrm{MHz}$). 
Details on isotope shifts and relative transition strengths of  ${}^{87}\mathrm{Sr}$ can be found in Ref.~\cite{S.Mauger2007}. 
The absorption profile of the atomic beam is then fitted using Eqs.~\eqref{eq:OD_summation} and \eqref{eq:OD_element}, with $n$ and $\sigma_t$ as fitting parameters (see Fig.~\ref{fig:absorption}). 
From the fitting results, the atomic beam density $n$ is obtained.

The atomic beam flux $\mathcal{F}$ ($\mathrm{atoms\,s^{-1}\,cm^{-2}}$) is given by
\begin{equation}
    \mathcal{F} = n v_{\text{beam}}, \label{eq:beam_flux} 
\end{equation}
where $v_{\text{beam}}$ is the mean speed of an effusive atomic beam~\cite{H.Metcalf1999}
\begin{equation}
    v_{\text{beam}} = \sqrt{\frac{9\pi k_B T_{\text{oven}}}{8m_{\mathrm{Sr}}}}, \label{eq:thermal_velocity}
\end{equation}
with $T_{\text{oven}}$ denoting the temperature of the atomic vapor in the oven.
Using the atomic beam cross-sectional area $A=\pi (l_{\text{int}}/2)^2$, the atomic beam intensity $J$ ($\mathrm{atoms\,s^{-1}\,sr^{-1}}$) is written as
\begin{equation}
    J = \frac{\mathcal{F}A}{\Omega_{\text{aperture}}}, \label{eq:flux_solid}
\end{equation}
where $\Omega_{\text{aperture}}$ is the solid angle subtended by the aperture as seen from the oven. For small divergence angles, it can be approximated by
\begin{equation}
    \Omega_{\text{aperture}} = \frac{A}{L_\text{aperture}^2}, \label{eq:solid_angle}
\end{equation}
with $L_{\text{aperture}}$ the distance from the oven to the aperture ($140\,\mathrm{mm}$).

On the other hand, the theoretical expression for the atomic beam intensity $J$ in the limit of no collisions occurring in the capillary tubes (Knudsen regime) is written as~\cite{M.Schioppo2012}
\begin{equation}
    J = \frac{\left(\pi a^2\right) \bar{v} n_{\text{oven}}}{4 \pi} N_{\text{cap}}, \label{eq:J_theory}
\end{equation}
where $\bar{v}=\sqrt{8 k_B T_{\text{oven}}/\pi m_{\text{Sr}}}$ is the mean atomic speed in the oven, $a$ is the capillary inner radius ($2a = 0.3\,\mathrm{mm}$),  $N_{\text{cap}}=130$ is the number of capillaries, and $n_{\text{oven}}$ is the atomic density of Sr in the oven. 
The atomic density is given by $n_{\text{oven}} = P/(k_B T_{\text{oven}})$, where $P$ is the Sr vapor pressure~\cite{L.R.Lide2003}:
\begin{align}
    \log_{10}\left(P/\mathrm{Pa}\right) = 14.232 - \frac{8572}{(T_{\text{oven}}/\mathrm{K})}  \nonumber \\
    -1.1926 \log_{10}\left(T_{\text{oven}}/\mathrm{K} \right). \label{eq:vapor_pressure}
\end{align}
A comparison between the experimental results and the theoretical predictions is shown in Fig.~\ref{fig:apparatus_results}.

In the collision-free regime the total flow rate of atoms emitted by the oven at $T_{\text{oven}} = 400\,\mathrm{{}^\circ C}$ is 
\begin{equation}
    \frac{dN_{\text{oven}}}{dt} =\frac{(\pi a^2) \bar{v} n_{\text{oven}} N_{\text{cap}} W}{4} \simeq 5.3 \times 10^{13} \mathrm{atoms/s}, \label{eq:flowrate}
\end{equation}
where $W = (8a/3L)/(1+8a/3L) \sim0.04$ is the transmission probability (Clausing factor) of the capillaries, with $L=10\,\mathrm{mm}$ being the capillary length.

\section{Loading rate of atoms into the MOT}
We describe the theoretical expression for the loading rate of atoms into the MOT. 
The loading rate $R$ can be written as
\begin{equation}
    R = J \Omega_{\text{trap}} f\left(v \le v_{c}\right), \label{eq:loading_rate}
\end{equation}
where the atomic beam flux $J$ is given by Eq.~(\ref{eq:J_theory}). 
The solid angle of the trapping region as viewed from the oven, $\Omega_{\text{trap}}$, is expressed as
\begin{equation}
    \Omega_{\text{trap}} = \frac{\pi r_{\text{trap}}^2}{L_{\text{trap}}^2}, \label{eq:trap_solid}
\end{equation}
where $L_{\text{trap}}$ is the distance from the oven to the trapping region ($350\,\mathrm{mm}$), and $r_{\text{trap}}$ is the radius of the trapping region ($18\,\mathrm{mm}/2 = 9\,\mathrm{mm}$). 
The factor $f(v \le v_{c})$ represents the fraction of atoms in the beam with velocity $v$ below the MOT capture velocity $v_c$, and is given by
\begin{equation}
    f\left(v \le v_{c}\right) = \frac{1}{2}  \left(\frac{m_{\mathrm{Sr}}}{k_B T_{\text{oven}}}\right)^2 \int_{0}^{v_c} v^3 \exp\left(-\frac{m_{\mathrm{Sr}} v^2}{2 k_B T_{\text{oven}}}\right) dv, \label{eq:beam_fraction}
\end{equation}
where $T_{\text{oven}}$ is the temperature of the Sr vapor inside the oven, and $m_{\mathrm{Sr}}$ is the mass of Sr.


When $v_c \ll \sqrt{k_B T_{\text{oven}}/m_{\mathrm{Sr}}}$, Eq.~\eqref{eq:beam_fraction} can be approximated as
\begin{align}
  f\left(v \le v_{c}\right) &\sim \frac{1}{2} \left(\frac{m_{\mathrm{Sr}}}{k_B T_{\text{oven}}}\right)^2 \int_{0}^{v_c} v^3 dv \nonumber \\ 
  &= \frac{1}{8} \left(\frac{m_{\mathrm{Sr}}}{k_B T_{\text{oven}}}\right)^2 {v_c}^4 \nonumber \\
  &\propto {v_c}^4. \label{eq:v4}
\end{align}
The loading rate is thus approximately proportional to the fourth power of the capture velocity.

\section{Dependence of the trapped atom number on the loading rate}
When repumping lights for the atoms in the $5s5p\,{}^3P_2$ and ${}^3P_0$ states are applied, the rate equation for the trapped atom number $N$ can be expressed as
\begin{equation}
    \frac{dN}{dt} = R - \gamma N - \tilde{\beta} N^2, \label{eq:rate_general}
\end{equation}
where $\gamma$ is the loss rate due to background gas collisions, and $\tilde{\beta}$ is the effective two-body loss coefficient ($\tilde{\beta}N$ is the effective per-atom loss rate due to two-body collisions). 
Here, we consider the steady-state trapped atom number $N_0$ in two limiting cases: (i) when background gas collisions dominate, and (ii) when two-body collisions dominate.

In case (i), $N_0$ can be expressed as
\begin{align}
    R - \gamma N_0 = 0, \nonumber \\
    N_0 = \frac{R}{\gamma}. \label{eq:background_steady}
\end{align}
Thus, $N_0$ is proportional to the loading rate of atoms into the MOT.

In case (ii), $N_0$ can be expressed as
\begin{align}
    R - \tilde{\beta} N_0^2 = 0, \nonumber \\
    N_0 = \sqrt{\frac{R}{\tilde{\beta}}}. \label{eq:twobody_steady}
\end{align}
Thus, $N_0$ is proportional to the square root of the loading rate of atoms into the MOT. The fitting result of Fig.~\ref{fig:loading_dual} yields $\tilde{\beta}=1.11(3) \times 10^{-7}\,\mathrm{s^{-1}}$. 


According to Ref.~\cite{T.Dinneen1999}, in the constant-volume regime, the density-normalized two-body loss coefficient $\beta$ ($\mathrm{cm^3\,s^{-1}}$) is related to $\tilde{\beta}$ as
\begin{equation}
    \beta = \tilde{\beta} \left(\sqrt{2\pi} r_e\right)^3, \label{eq:twobody_coeff}
\end{equation}
where $r_e$ is the $1/e$ radius of the MOT density distribution, given by $n=n_0 \exp\!\left[-\left(r/r_e\right)^2\right]$, with $r$ being the distance from the MOT center and $n_0$ the central density. 



\bibliography{references.bib}

\end{document}